\documentclass{ws-procs9x6-cpt19}
\usepackage{amsmath}
\usepackage{url} 

\begin{document}
\newcommand{\refeq}[1]{(\ref{#1})}
\def\etal {{\it et al.}}
\def\Bperp {$B_\text{osc}\perp B_\text{ext}$}
\def\Bparr {$B_\text{osc}\parallel B_\text{ext}$}

\title{Progress Towards Ramsey Hyperfine Spectroscopy in ASACUSA}

\author{A.\ Nanda}

\address{Stefan Meyer Institute, Austrian Academy of Sciences,\\
Boltzmanngasse 3, 1090 Vienna, Austria}

\author{On behalf of the ASACUSA-CUSP Collaboration\footnote{\url{http://asacusa.web.cern.ch/ASACUSA/asacusaweb/antihydrogen_cusp/main.shtml}}}

\begin{abstract}
This proceedings contribution reports on progress 
on the design of a Ramsey-type spectrometer 
that will be used for the spectroscopy of 
the ground-state hyperfine structure 
of both hydrogen and deuterium. 
\end{abstract}

\bodymatter
\section{Antihydrogen spectroscopy program of ASACUSA}
The ASACUSA-CUSP 
(Atomic Spectroscopy And Collisions Using Slow Antiprotons) 
antihydrogen experiment 
based at the Antiproton Decelerator facility of CERN 
aims to perform precise tests of CPT symmetry 
by measuring the ground-state hyperfine structure (GS-HFS) of antihydrogen atoms 
and comparing it with that of hydrogen.\cite{WidmannHI} 
A brief account on the status of ASACUSA's antihydrogen program 
can be found within these proceedings.\citep{procMartin}  
The currently applied Rabi-type beam-spectroscopy method 
can be made more precise by the implementation of Ramsey's technique 
of using two separated oscillatory fields.\cite{Ramseypaper} 
This improvement will first be tested on hydrogen and later on deuterium, 
which can help constrain the associated SME coefficients.\cite{SMEpaper}

\section{Design of the spectrometer and its scope}
An in-beam spectroscopy method comprises a spin-selected atomic beam, 
an oscillating field to drive transitions between two different spin states, 
a field gradient to select spin states of interest, 
and a detection scheme to measure the amount of beam in this state. 

\begin{figure}[!ht]

  \begin{minipage}[b]{0.47\textwidth}
    \centering
    \includegraphics[width=\textwidth]{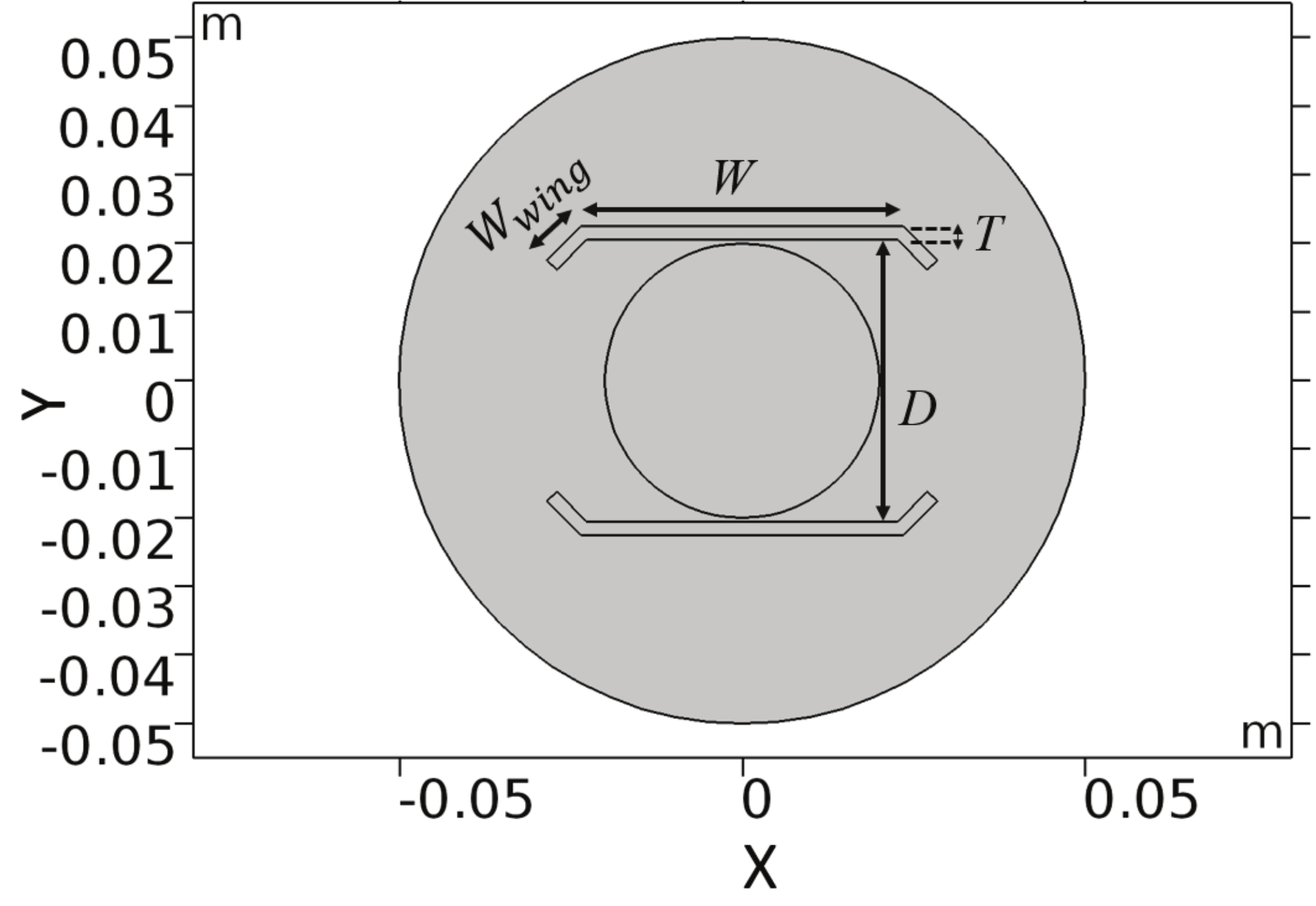}
    \caption{Cross-sectional view of the strip\-line device.}
    \label{ecylinder}
  \end{minipage}
  \hfill
  \begin{minipage}[b]{0.47\textwidth}
    \centering
    \includegraphics[width=\textwidth]{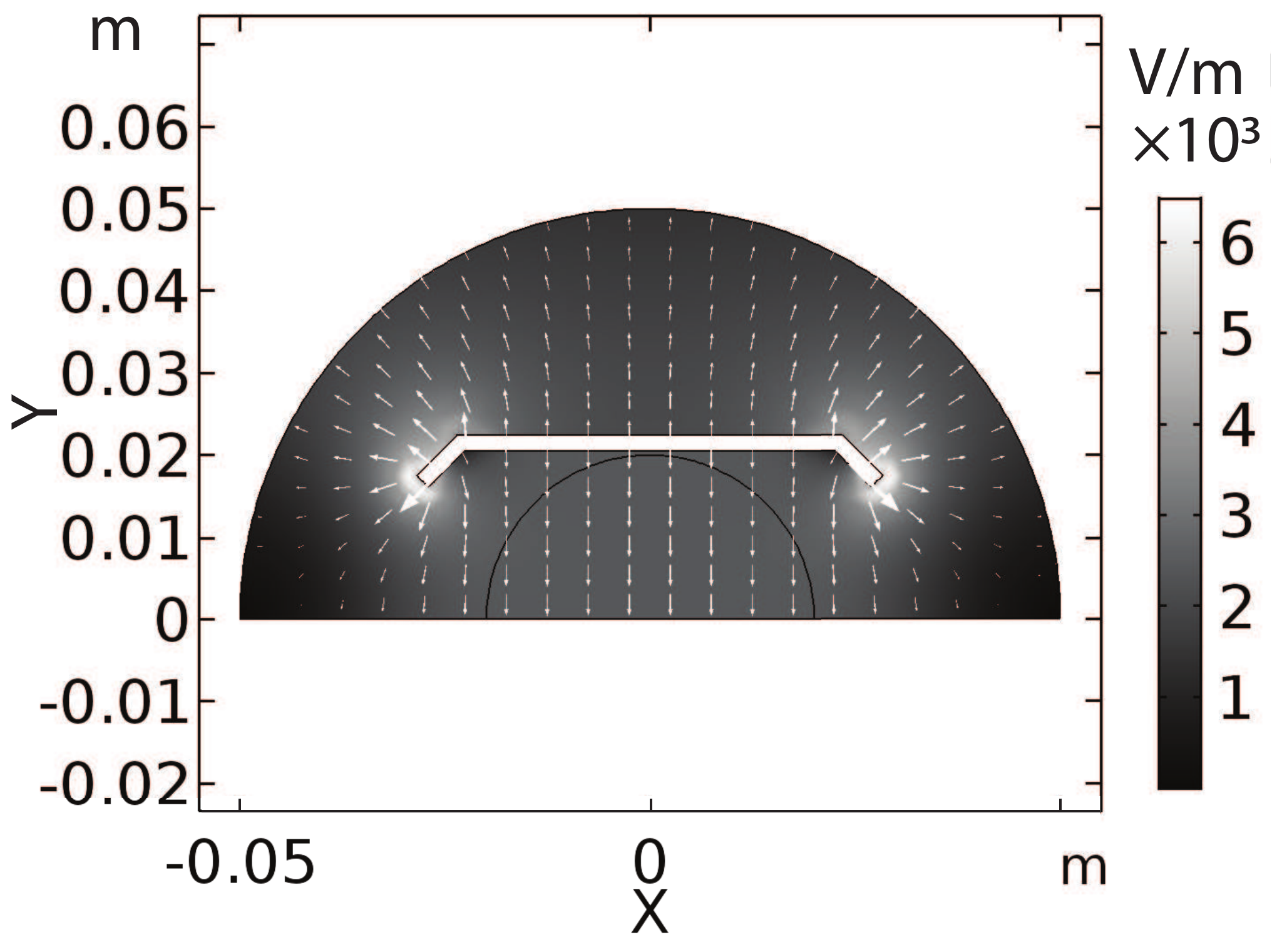}
    \caption{Electric field in the simulated structure.}
    \label{fig:efield}
  \end{minipage}
  
\end{figure}

In this section we report the progress 
on a new broadband microwave-generation system 
that is well suited for in-beam spectroscopy 
using the Ramsey method. 
Simulations for a stripline traveling-wave device 
have been studied with COMSOL 5.3 
using the 2D electrostatic study feature in the AC/DC Module.\cite{COMSOL} 
A cross-sectional view of the device 
consisting of two electrodes and housed inside a $100\,$mm diameter cylinder 
is shown in Fig.~\ref{ecylinder}. 
The $40\,$mm diameter circle in between the two electrodes 
represents the area where the hydrogen beam will be present. 
The electrostatic problem was solved for only half the structure 
and using the central plane between the two electrodes as a virtual ground. 
The potential at the electrode boundaries was $50\,$V 
while the remaining boundaries were all defined as ground. 
The resulting electric field is shown in Fig.~\ref{fig:efield}. 
The electric field is perpendicular to the electrode, 
hence when operated in transverse electromagnetic (TEM) mode, 
the oscillating magnetic field ($B_\text{osc}$) will be parallel to the electrode 
and perpendicular to the axis of the cylindrical housing. 
The impedance of the structure was calculated as $Z = 1/(c\cdot K)$, 
where $c$ is the speed of light in free space 
and $K$ is the computed capacitance. 
The electric-field inhomogeneity 
(ratio of the standard deviation to the average of the norm of the electric field) 
of the structure was studied in the area where the hydrogen beam would be present.
The geometrical configuration with 
$W = 47\,$mm, 
$D = 41\,$mm, 
$T = 2\,$mm, 
and $W_{wing} = 7\,$mm 
has an impedance of $50.037\,\Omega$, 
and $0.67\%$ electric-field inhomogeneity.

The $\sigma$ and $\pi$ transitions\cite{procMartin} 
require different orientations of $B_\text{osc}$ 
relative to an external static magnetic field ($B_\text{ext}$).\cite{procMartin} 
This means that with the stripline structure 
and an axial magnetic field along the beam direction, 
we can only measure the $\pi$ transition. 
To measure the $\sigma$ transition, 
a TE$_{110}$ mode cavity can be used, 
which has $B_\text{osc}$ along its axis. 
As the $\sigma$ transition in hydrogen is insensitive to Lorentz violation,\cite{SMEpaper} 
it can be used as a benchmark against the $\pi$ transitions. 

Due to the broadband nature of the TEM travelling waveguide 
we can measure the GS-HFS of deuterium around $330\,$MHz, 
in addition to that of hydrogen at $1.42\,$GHz. 
The energy shifts of the hyperfine Zeeman levels in deuterium 
due to the effects of CPT and Lorentz violation 
within the Standard-Model Extension framework 
are shown in Eq.~(123) of Ref.~\refcite{SMEpaper}, 
and they depend on the expectation values 
the momentum of the proton relative to the center of mass of the deuteron 
($\textbf{p}_{pd}$) 
and its higher powers ($\langle \textbf{p}^k_{pd} \rangle$).\cite{SMEpaper} 
This feature, 
being significantly different from that of hydrogen, 
leads to the  enhancement of the sensitivity to coefficients 
for Lorentz and CPT violation\cite{SMEpaper} by a factor of $10^9$ 
for coefficients with $k = 2$ 
and $10^{18}$ for coefficients with $k = 4$.

\section{Summary and outlook}
We have reported on the geometry of a stripline structure 
that can be used as an interaction zone 
in Ramsey-type spectroscopy for GS-HFS measurements of hydrogen and deuterium. 
The advantages of using two different microwave regions 
for the $\sigma$ and $\pi$ transitions have also been discussed.

Next, 
the transition from RF feedthroughs to the striplines 
for feeding microwaves into the device 
will be designed with input from further simulations. 
Later, 
possibilities for measuring $\sigma$ transitions in deuterium 
will be studied 
with simulations of split ring resonators\cite{Har1981} 
instead of a TE$_{110}$ cavity.

\section*{Acknowledgments}
This project is supported by 
the European Union’s Horizon 2020 research and innovation program 
under the Marie Skłodowska-Curie grant agreement No.\ 721559 
and the Austrian Science Fund FWF, 
Doctoral Program No.\ W1252-N27. 
We also acknowledge Dr.\ Fritz Caspers 
for helping us with the design of microwave devices.


\begin{thebibliography}{x}
\bibitem{WidmannHI}
E.\ Widmann \etal,
Hyperfine Interact.\ {\bf 240}, 5 (2019).

\bibitem{procMartin}
M.C.\ Simon,
these proceedings (also: arXiv:1910.03959).

\bibitem{Ramseypaper}
N.F.\ Ramsey \etal,
Rev.\ Mod.\ Phys.\ {\bf 62}, 541 (1990).

\bibitem{SMEpaper}
V.A.\ Kosteleck\'y \etal,
Phys.\ Rev.\ D {\bf 92}, 056002 (2015).

\bibitem{COMSOL}
AC/DC Module User's Guide, 
Version 5.3, 
COMSOL, 
Inc., 
www.comsol.com.

\bibitem{Har1981}
W.N.\ Hardy \etal,
Rev.\ Sci.\ Instrum.\ {\bf 52}, 213 (1981).

\end{thebibliography}
\end{document}